\documentclass[12pt]{article}
\usepackage{amssymb}
\usepackage{graphicx}
\usepackage{subfigure}
\usepackage{psfrag}

\newcommand{\beq}{\begin{equation}}
\newcommand{\eeq}{\end{equation}}
\newcommand{\bea}{\begin{eqnarray}}
\newcommand{\eea}{\end{eqnarray}}
\newcommand{\ba}{\begin{array}}
\newcommand{\ea}{\end{array}}
\newcommand{\bi}{\begin{itemize}}
\newcommand{\ei}{\end{itemize}}
\newcommand{\bn}{\begin{enumerate}}
\newcommand{\en}{\end{enumerate}}
\newcommand{\bc}{\begin{center}}
\newcommand{\ec}{\end{center}}
\renewcommand{\l}{\left}
\renewcommand{\r}{\right}

\newcommand{\ol}{\overline}

\newcommand{\be}{\beta}
\newcommand{\ga}{\gamma}
\newcommand{\de}{\delta}

\newcommand{\nl}{\nonumber\\}

\newcommand{\wt}[1]{\widetilde{#1}}

\begin{document}
\tolerance=100000


\vspace*{\fill}

\begin{center}
{\Large \bf  A solution to $B \to \pi \pi$ puzzle and $B \to K K$}\\[3.cm]

{\large\bf Seungwon Baek\footnote{sbaek@korea.ac.kr}}
\\[7mm]

{\it
The Institute of Basic Science, and Department of Physics, Korea University, Seoul 136-701, Korea
}\\[10mm]
\end{center}

\vspace*{\fill}

\begin{abstract}
{\small\noindent
The large ratio of color-suppressed tree amplitude to color-allowed one in $B \to \pi \pi$ decays
is difficult to understand within the Standard Model, which is known as the ``$B \to \pi\pi$ puzzle".
The two tree diagrams contain the up- and charm-quark component of penguin amplitude, $P_{uc}$,
which cannot be separated by measuring $B \to \pi \pi$ decays alone. We show that the measurements of
the branching ratio and direct CP asymmetry of $B^+ \to K^+ \overline{K^0}$ decay enable one to
disentangle the $P_{uc}$ with two-fold ambiguity. One of the two degenerate solutions of the $P_{uc}$
can solve the $B \to \pi \pi$ puzzle by giving $|C/T|\sim 0.3$ which is consistent with the expectation
in the Standard Model.
We also show that the two solutions can be discriminated by the measurement of the indirect
CP-asymmetry of $B^0 \to K^0 \ol{K^0}$. We point out that if the $B \to \pi \pi$ puzzle is solved
in this way, the corresponding puzzle in $B \to \pi K$ decays should have a different
origin.
}
\end{abstract}

\vspace*{\fill}

\newpage

\section{Introduction}
In the Standard Model (SM), rare charmless nonleptonic $B \to \pi\pi$ decays
provide valuable information on the inner angles of
the unitarity triangle of Cabbibo-Kobayashi-Maskawa (CKM) matrix~\cite{B2pipi}.
In these decays, dominated by $b \to d q \ol{q}$ ($q=u,d$) at the quark level,
one can measure the angle $\alpha$ through
the isospin analysis~\cite{London-Gronau-1990}.
In addition, new physics (NP) beyond the SM
can affect the decay processes~\cite{Baek:pipi}.

In the diagrammatic approach~\cite{GHLR}, the $B \to \pi \pi$ decay amplitudes are
dominated by color-allowed (suppressed) tree diagram $T(C)$ and QCD-penguin
diagram $P$. These diagrammatic amplitudes are expected to be hierarchical in size,
{\it i.e.}, $|C| \approx |P|$ much less than $|T|$.
However, the experimental data of the branching ratio (BR)
of $B \to \pi^0 \pi^0$ decay requires a large ratio of the color-suppressed tree
diagram to the color-allowed one, which
is in contradiction to the hierarchy. This is known as the ``$B \to \pi \pi$-puzzle''.

Recently another $B$-decay modes, $B^+ \to K^+ \ol{K^0}$ and $B^0 \to K^0 \ol{K^0}$,
which have identical $b \to d$ transitions at the quark
level have been observed
by the BaBar~\cite{Babar:B2KK} and Belle~\cite{Belle:B2KK} Collaborations.
They have also measured the time-dependent CP asymmetries, ${\cal A}_f$
 and ${\cal S}_f$ defined by
\bea
 \frac{\Gamma_{\ol{B^0}\to f}(\Delta t)- \Gamma_{{B^0}\to f}(\Delta t)}
 {\Gamma_{\ol{B^0}\to f}(\Delta t)+ \Gamma_{{B^0}\to f}(\Delta t)} \equiv
 {\cal A}_f \cos(\Delta m \Delta t) + {\cal S}_f \sin(\Delta m \Delta t).
\eea
We study the implications
of these newly discovered channels on the extraction of the SM amplitudes.
Other studies on $B \to KK$ can be found in~\cite{B2KK}.

We take
{\it i)} $SU(3)$ flavor symmetry of strong interactions and
{\it ii)} smallness of annihilation and exchange topologies
as working assumptions.
Then  $B^+ \to K^+ \ol{K^0}$ and $B^0 \to K^0 \ol{K^0}$ decays become pure penguin modes.

The penguin amplitude can be decomposed into two parts depending on
the quarks running inside the loop, up- and charm-quark penguin $P_{uc}$
and top- and charm-quark penguin $P_{tc}$.
In $B \to \pi\pi$ decays  $P_{uc}$ can always be absorbed into $T$ and $C$.
As a consequence, it is impossible to extract pure $T$ or $C$
amplitudes from $B \to \pi \pi$ measurements alone.

We show that when combining the $B \to \pi \pi$ and $ B \to K K$ decays, we can extract
the tree amplitudes and $P_{uc}$ separately. So we can test the ratio $|C/T|$
 without the contamination of $P_{uc}$ penguin in the $b\to q\ol{q}$ transitions.

We will demonstrate that a solution obtained from the $B \to \pi \pi$ data,
${\cal B}(B^+ \to K^+ \ol{K^0})$,
and ${\cal A}_{\rm CP}(B^+ \to K^+ \ol{K^0})$
has $|C/T| \sim 0.3$ which is
in accord with the SM hierarchy. First, we use the analytic expressions for
the BR's and CP asymmetries to show that all the relevant amplitudes,
$T$, $C$, $P_{tc}$, $P_{uc}$, and the angle $\ga$ of the unitarity triangle
can be extracted with two-fold ambiguity.

Then we get the diagrammatic amplitudes numerically  by
performing a $\chi^2$-fit to the current data.
Discarding the indirect CP
asymmetry of $B \to K^0  \ol{K^0}$, ${\cal S}_{\rm CP}(B \to K^0  \ol{K^0})$, which has huge errors,
we obtain two degenerate solutions which minimize $\chi^2$. One solution, as mentioned above,
is acceptable.
However, the other solution gives $|C/T| \sim 1.3$ which is unsatisfactory.

From the two solutions we can predict ${\cal S}_{\rm CP}(B \to K^0  \ol{K^0})$. The favored solution
with $|C/T| \sim 0.3$ gives large positive ${\cal S}_{\rm CP}(B \to K^0  \ol{K^0})$
while the other solution which looks unphysical predicts large
negative ${\cal S}_{\rm CP}(B \to K^0  \ol{K^0})$.
Therefore if the ${\cal S}_{\rm CP}(B \to K^0  \ol{K^0})$ is measured more accurately in the near future,
we can discriminate the two solutions and confirm the physical solution.

It is also well known that   $|C'/T'| \sim 1$ in $B \to \pi K$ decays,
which is again difficult to understand in the SM framework~\cite{B2piK_puzzle}.
In $B \to \pi K$ decays, the contamination of $P'_{uc}$ does not play
an important role because it is suppressed by CKM factors. Therefore the ``$B \to \pi K$
puzzle'', if it remains in the future experimental data, should be solved
by other mechanisms such as new physics effects.

This paper is organized as follows.
In section~\ref{sec:B2pipi},  using the current experimental $B \to \pi \pi$
data only, we find that the $B \to \pi \pi$ puzzle  is still there.
In section~\ref{sec:pipiKK}, we confirm the possibility that the $B \to \pi \pi$ puzzle
is simply the effect of $P_{uc}$ and there is actually no $B \to \pi \pi$ puzzle
in the ``bare" ratio of $|C/T|$. We conclude in section~\ref{sec:conclusion}.

\section{The current status of $B \to \pi \pi$ puzzle}
\label{sec:B2pipi}

The diagrammatic amplitudes provide a useful
parametrization for nonleptonic $B$-meson decay processes
 because it is independent of theoretical models for the calculation
of hadronic matrix elements~\cite{GHLR}.
The decay amplitudes of $B\to \pi\pi$'s are written as
\bea
\sqrt{2} A(B^+ \to \pi^+ \pi^0) &=& -\l(T + C + P_{\rm EW} + P^C_{\rm EW} \r),
\nl
A(B^0 \to \pi^+ \pi^-) &=& -\l(T + P + {2 \over 3} P^C_{\rm EW} + E + PA \r), \nl
\sqrt{2} A(B^0 \to \pi^0 \pi^0) &=& -\l(C - P +P_{\rm EW}
+ {1 \over 3}
P^C_{\rm EW} - E -PA\r).
\label{eq:pipi_full}
\eea
Here $T$, $C$, $P$, $P_{\rm EW}^{(C)}$, $E$, and $PA$ represent
the tree, color-suppressed tree, QCD-penguin, (color-suppressed)
electroweak-penguin,
exchange, and penguin-annihilation amplitudes, respectively.
In general, the diagrammatic amplitudes may have a weak phase as well as
a strong phase.

We can further decompose the QCD penguin diagram $P$ which depends on the
quarks running inside the loop, as
\bea
P &=& V_{ud} V^*_{ub} P_u + V_{cd} V^*_{cb} P_c + V_{td} V^*_{tb} P_t \nl
  &=& V_{ud} V^*_{ub} (P_u-P_c) + V_{td} V^*_{tb} (P_t-P_c) \nl
  &\equiv& P_{uc} e^{i\gamma} + P_{tc} e^{-i\beta},
\eea
where we use the unitarity property of CKM matrix
and explicitly write the weak phase dependence for the amplitudes.
The relative sizes of the diagrammatic amplitudes are expected to have the following
hierarchical structure:
\bea
\begin{array}{ccccc}
\hline
O(1) & O(\bar{\lambda}) & O(\bar{\lambda}^2) & O(\bar{\lambda}^3) & O(\bar{\lambda}^4) \\
\hline
 |T|  &  |C|,|P|  &  |P_{\rm EW}|  &  |P^C_{\rm EW}|  &  |E|,|PA| \\
\hline
\label{eq:hierarchy}
\end{array}
\eea
where $\bar{\lambda}$ is expected to be of order of $0.2$ -- $0.3$.
We neglect terms of order $\bar\lambda^2$ or higher and thus, consider only $T, C,$ and $P$.

Taking into account the decay amplitudes containing terms up to $O(\bar{\lambda})$,
we find that the amplitudes in $B \to \pi \pi$ decays are given by
\bea
-\sqrt{2} A(B^+ \to \pi^+ \pi^0) &=& \l(T + C \r)  e^{i\gamma},
\nl
-A(B^0 \to \pi^+ \pi^-) &=& T e^{i\gamma} + P_{tc} e^{-i\beta} + P_{uc} e^{i\gamma} , \nl
-\sqrt{2} A(B^0 \to \pi^0 \pi^0) &=& C^{i\gamma} - P_{tc} e^{-i\beta}- P_{uc} e^{i\gamma} ,
\label{eq:all}
\eea
where the weak-phase dependence is explicitly written. From the above expression
we notice that the $P_{uc}$ term can always be absorbed into tree
diagrams $T$ and $C$ by the redefinition
\bea
  \wt{T} &=& T + P_{uc}, \nl
  \wt{C} &=& C - P_{uc}.
\eea
Then the $B \to \pi \pi$ decay amplitudes are further
simplified:
\bea
-\sqrt{2} A(B^+ \to \pi^+ \pi^0) &=& \l(\wt{T} + \wt{C} \r)  e^{i\gamma},
\nl
-A(B^0 \to \pi^+ \pi^-) &=& \wt{T} e^{i\gamma} + P_{tc} e^{-i\beta} , \nl
-\sqrt{2} A(B^0 \to \pi^0 \pi^0) &=& \wt{C} e^{i\gamma} - P_{tc} e^{-i\beta}.
\label{eq:B2pipi_only}
\eea
As a consequence, the $P_{uc}$ term cannot be detected if
we use the $B \to \pi\pi$ data only.

\begin{table}[t]
\center
\begin{tabular}{cccc}
\hline
\hline
Mode & ${\cal B}[10^{-6}]$ & ${\cal A}_{\rm CP}$ & ${\cal S}_{\rm CP}$ \\ \hline
$B^+ \to \pi^+ \pi^0$ & $5.7 \pm 0.4$ & $0.04 \pm 0.05$ & \\
$B^0 \to \pi^+ \pi^-$ & $5.16 \pm 0.22$ & $0.38 \pm 0.07$ & $-0.61 \pm 0.08$ \\
$B^0 \to \pi^0 \pi^0$ & $1.31 \pm 0.21$ & $0.36 \pm 0.33$ & \\
\hline
$B^+ \to K^+ \ol{K^0}$ & $1.36^{+0.29}_{-0.27}$ & $0.12^{+0.17}_{-0.18}$ & \\
$B^0 \to K^0 \ol{K^0}$ & $0.96^{+0.21}_{-0.19}$ &
$^{-0.58^{+0.73}_{-0.66}\pm 0.04\;\;(\rm Belle)}_{\phantom{-}0.40\pm 0.41 \pm 0.06 \;(\rm BaBar)}$
& $-1.28^{+0.80\,+0.11}_{-0.73\,-0.16}$\\
\hline
\hline
\end{tabular}
\caption{The current experimental data for CP averaged branching ratios ($BR$),
direct CP asymmetries (${\cal A}_{\rm CP}$)
and indirect CP asymmetries (${\cal S}_{\rm CP}$)
of $B \to \pi \pi$ and $B \to \pi K$ decays~\cite{Babar:B2KK,Belle:B2KK,HFAG}.}
\label{tab:data}
\end{table}

The current experimental data for the CP-averaged branching ratio (${\cal B}$),
direct CP-asymmetry (${\cal A}_{\rm CP}$), and indirect CP-asymmetry
(${\cal S}_{\rm CP}$) are shown in Table~\ref{tab:data}.
If we take the weak phase $\be=21.2^\circ$ measured through the $b \to c \ol{c} s$
transition, we have 6 parameters in (\ref{eq:B2pipi_only}).
We fit 6 parameters in (\ref{eq:B2pipi_only}) to
the 7 measurements. The results are shown in Table~\ref{tab:B2pipi}.
The $\chi^2_{\rm min}/{\rm d.o.f}=0.81/1$ is
very good.
The angle $\ga=(65.5 \pm 14.1)^\circ$ (all the angles in this paper are
in the unit of degrees) is consistent with the global CKM fit value
$\ga = 55.6^{+4.1}_{-2.5}$ or  $68.0^{+2.6}_{-7.0}$ obtained by CKMfitter group~\cite{CKMfitter}.
However, we obtain
\bea
\frac{|\wt{C}|}{|\wt{T}|} = 0.71 \pm 0.18,
\label{eq:CoverT}
\eea
which is larger than the expectation (\ref{eq:hierarchy}) by about factor 3.
This large ratio is mainly due to the unexpectedly large ${\cal B}(B\to \pi^0 \pi^0)$.
We confirm the $B\to \pi \pi$ puzzle~\cite{Li_Mishima06} is still there.

\begin{table}[tb]
\begin{center}
\begin{tabular}{cccc}
\hline
\hline
$\gamma$ & $(|\wt{T}|,\delta_{\wt{T}})$  & $(|\wt{C}|,\delta_{\wt{C}})$  &  $|P_{tc}|$ \\
\hline
$65.5 \pm 14.1$ & $(22.4 \pm 0.7, 36.0 \pm 22.6)$ & $(15.8 \pm 4.1, -16.5 \pm 12.7)$ &
$8.33 \pm 4.34$\\
\hline
\hline
\end{tabular}
\end{center}
\caption{The results for the fit to the $B\to \pi\pi$ data alone. We obtained
$\chi^2_{\rm min}/{\rm d.o.f}=0.81/1$.
The strong phase of $P_{tc}$ is set to zero.  The magnitudes and angles are
in the unit of $eV$'s and degrees, respectively.}
\label{tab:B2pipi}
\end{table}

\section{Combining $B \to \pi \pi$ with $B \to K K$}
\label{sec:pipiKK}

Since from the $B\to \pi \pi$ data alone we cannot disentangle the $P_{uc}$,
we include the $B \to K K$ data in the analysis.
There are three decay modes for these $b \to d(u) s \ol{s}$ transitions:
\bea
A(B^+ \to K^+ \ol{K^0}) &=& P + A -{1 \over 3} P^C_{\rm EW}, \nl
A(B^0 \to K^0 \ol{K^0}) &=& P + PA -{1 \over 3} P^C_{\rm EW}, \nl
A(B^0 \to K^+ K^-) &=& -E -PA.
\eea
When we neglect small contributions from annihilation and electroweak penguins, we simply get
\bea
A(B^+ \to K^+ \ol{K^0}) &=& A(B^0 \to K^0 \ol{K^0}) =P = P_{tc} e^{-i \be}(1+ r e^{i \de_{uc}} e^{i(\be+\ga)}),
\label{eq:amp_KK}
\eea
where $r\equiv|P_{uc}/P_{tc}|$ and $\de_{uc}$ is the strong phase of $P_{uc}$ (We set $\de_{tc} \equiv 0$.).
The $A(B^0 \to K^+ K^-)$ is negligible.
Therefore we expect {\it i)} ${\cal B}(B^0 \to K^+ K^-)$ is very small,
which is confirmed by the experimental data~\cite{Babar:B2KK,Belle:B2KK},
{\it ii)} ${\cal B}(B^+ \to K^+ \ol{K^0}) \approx {\cal B}(B^0 \to K^0 \ol{K^0})$,
which agrees with the data, and {\it iii)}
${\cal A}_{\rm CP}(B^+ \to K^+ \ol{K^0}) \approx {\cal A}_{\rm CP}(B^0 \to K^0 \ol{K^0})$.
For the last relation, the results of BaBar and Belle
are contradictory with each other and we need to wait for more data
to confirm or reject it.
As one can see in Table~\ref{tab:data}, the indirect CP asymmetry
${\cal S}_{\rm CP}(B^0 \to K^0 \ol{K^0})$ has still large experimental errors.
For these reasons we use only ${\cal B}(B^+ \to K^+ \ol{K^0})$, ${\cal B}(B^0 \to K^0 \ol{K^0})$,
and ${\cal A}_{\rm CP}(B^+ \to K^+ \ol{K^0})$ for the numerical analysis which will
be presented below.

First, let us demonstrate that the inclusion of just two more data,
${\cal B}(B^{+} \to K^{+} \ol{K^0})$ and ${\cal A}_{\rm CP}(B^+ \to K^+ \ol{K^0})$,
is enough to extract $P_{uc}$.
Given ${\cal B}(B^{+} \to K^{+} \ol{K^0})$, one can obtain
\bea
 R \equiv \frac{|A(B^{+} \to K^{+} \ol{K^0})|^2+|A(B^{-} \to K^{-} K^0)|^2}
 {2|P_{tc}|^2} = 1+r^2+2 r \cos(\beta+\gamma) \cos \delta_{uc}
 \label{eq:Ampsq}
\eea
for the ratio of average amplitude squared to $|P_{tc}|^2$.
The information on $|P_{tc}|$ is obtained from $B \to \pi \pi$.

\begin{figure}[t]
\begin{center}
\psfrag{rrr}{$r$}
\psfrag{ddd}{$\delta_{uc}$}
\subfigure[]{
\includegraphics[width=0.45\textwidth]{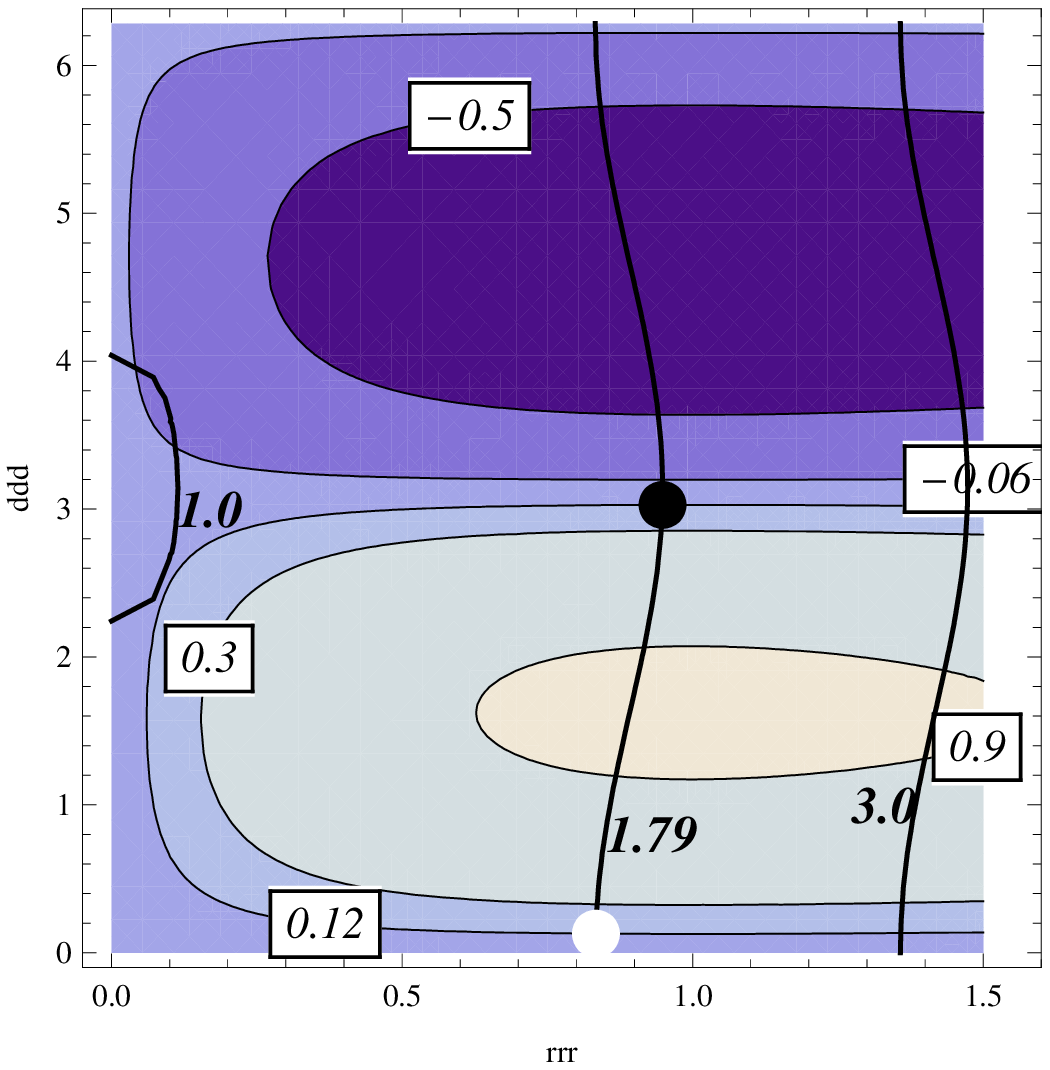}
\label{fig:CPA_a}
}
\subfigure[]{
\includegraphics[width=0.45\textwidth]{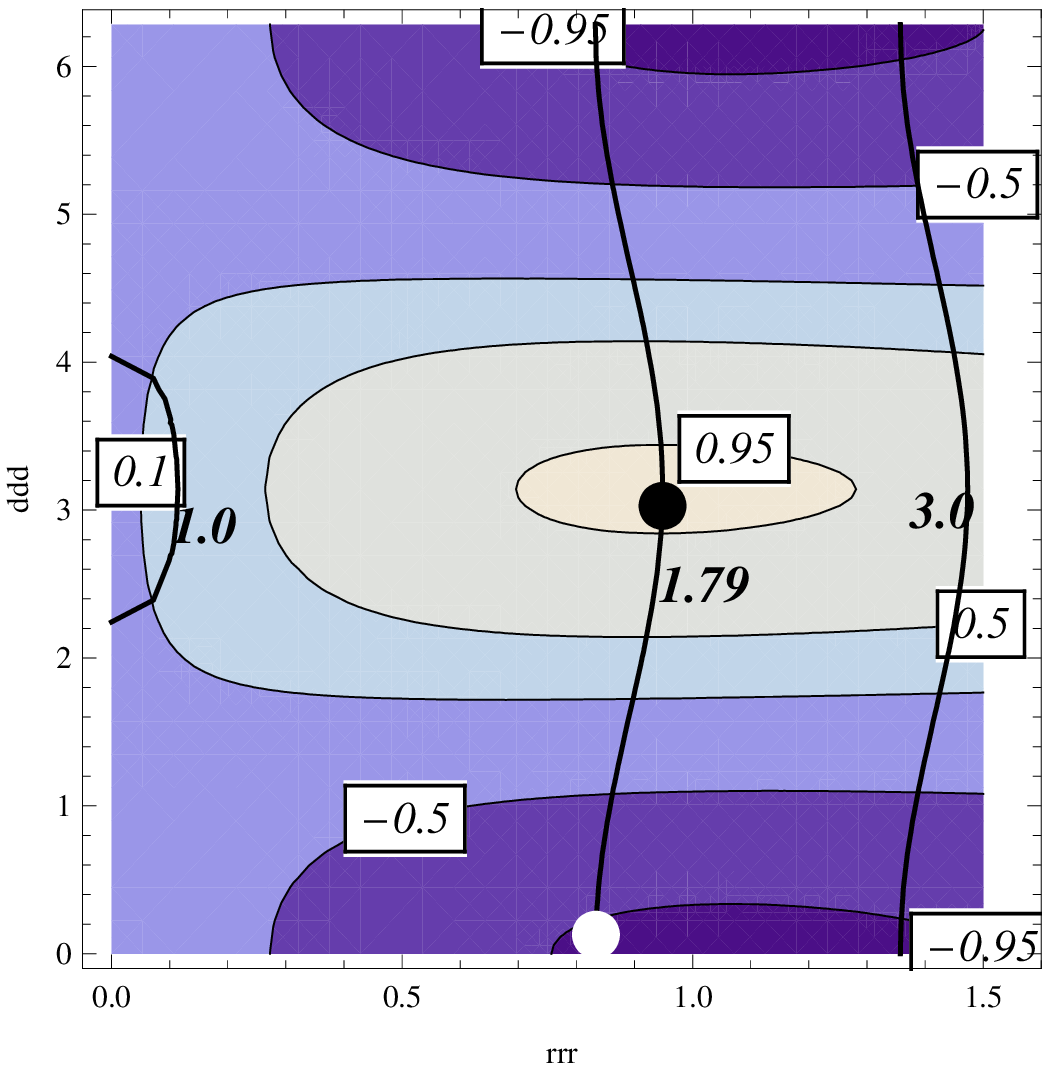}
\label{fig:CPA_b}
}
\end{center}
\caption{Contour plots for constant(a) direct and (b) indirect CP asymmetries as
a function of $r=|P_{uc}/P_{tc}|$ and the strong phase $\de_{uc}$ (Filled contours).
Also contours for constant $R$ are shown on each plot.
The lines for ${\cal A}_{\rm CP}(B^{+(0)} \to K^{+(0)} \ol{K^0})=0.12$ and
$R=1.79$ which correspond to the experimental central values
cross at two points in (a), denoted by black and white dots.
The points are also shown in (b).
}
\label{fig:CPA}
\end{figure}

The direct and indirect CP asymmetries for $B^{+(0)} \to K^{+(0)} \ol{K^0}$ are given by
\bea
 {\cal A}_{\rm CP}(B^{+(0)} \to K^{+(0)} \ol{K^0}) &=& \frac{2 r \sin(\beta+\gamma) \sin\de_{uc}}
  {1+r^2+2 r \cos(\beta+\gamma) \cos \delta_{uc}}, \nl
 {\cal S}_{\rm CP}(B^0 \to K^0 \ol{K^0}) &=& -\frac{r^2 \sin 2(\beta+\gamma)+2 r \sin(\be+\ga) \cos\de_{uc}}
 {1+r^2+2 r \cos(\be+\ga) \cos \de_{uc}},
 \label{eq:CnS}
\eea
respectively.
When the angles $\beta$ and $\gamma$ are given, $R$ and the CP asymmetries are functions of
$r$ and $\de_{uc}$ only.
From (\ref{eq:Ampsq}) and (\ref{eq:CnS}), we can determine $r$ and $\de_{uc}$ with $|P_{tc}|$ and $\ga$
determined from $B \to \pi \pi$ decays.

In Figure~\ref{fig:CPA}, we show filled contour plots of
${\cal A}_{\rm CP}(B^{+(0)} \to K^{+(0)} \ol{K^0})$  (Figure~\ref{fig:CPA}(a)) and
${\cal S}_{\rm CP}(B^0 \to K^0 \ol{K^0})$ (Figure~\ref{fig:CPA}(b)).
We also draw contours for various values of $R$ with thick lines on each plot.
For these plots the CP phases are taken to be $\beta=21.2^\circ$ and $\gamma=65.5^\circ$.

In Figure~\ref{fig:CPA}(a), the lines for ${\cal A}_{\rm CP}(B^{+(0)} \to K^{+(0)} \ol{K^0})=0.12$ and
$R=1.79$ which correspond to the experimental central values
cross at two points denoted by the black and white dots.
Thereby we get $|P_{uc}|$ and $\de_{uc}$ with two-fold ambiguity
once $|P_{tc}|$ is obtained from the $B\to \pi \pi$ data.

In Figure~\ref{fig:CPA}(b), we can see that the two solutions obtained from
Figure~\ref{fig:CPA}(a) predict very distinct ${\cal S}_{\rm CP}(B^0 \to K^0 \ol{K^0})$.
Therefore we can discriminate the two solutions by measuring
${\cal S}_{\rm CP}(B^0 \to K^0 \ol{K^0})$ in near future.
From the figure we note that the prediction for ${\cal S}_{\rm CP}(B^0 \to K^0 \ol{K^0})$,
especially its sign, is almost insensitive to the change of $r$
unless $r$ is too small.

Now, we proceed to obtain numerical solutions.
For this purpose we perform a $\chi^2$-fit to the seven $B\to \pi \pi$ data,
${\cal B}(B^{+(0)} \to K^{+(0)} \ol{K^0})$, and ${\cal A}_{\rm CP}(B^+ \to K^+ \ol{K^0})$.
We do not include the CP asymmetries of $B^0 \to K^0 \ol{K^0}$ to the fit because
they have huge errors and/or BaBar and Belle data are not consistent with each other.

\begin{table}[tb]
\begin{center}
\begin{tabular}{ccc}
\hline
\hline
 $\ga$ & $(|T|,\de_T)$  & $(|C|,\de_C)$  \\
\hline
$65.5 \pm 14.1$ & $(28.7 \pm 4.8, 25.3 \pm 11.6)$ & $(8.14 \pm 3.19, -26.12 \pm 30.6)$ \\
$65.5 \pm 14.1$ & $(16.7 \pm 1.7, 25.3 \pm 11.6)$ & $(22.3 \pm 7.1, -9.25 \pm 9.50)$ \\
\hline
\hline
$(|P_{tc}|,\de_{tc})$ & $(|P_{uc}|,\de_{uc})$ & $|C/T|$\\
\hline
$(8.33 \pm 4.33, 0)$ & $(7.89 \pm 6.82, 174 \pm 10)$ & $0.28 \pm 0.15$ \\
$(8.33 \pm 4.34, 0)$ & $(6.93 \pm 3.58, 7.42 \pm 11.2)$ & $1.34 \pm 0.50$\\
\hline
\hline
\end{tabular}
\end{center}
\caption{The results for the fit to the $B\to \pi\pi$ and $B\to KK$ data.
We obtained $\chi^2_{\rm min}=0.81$.
The strong phase of $P_{tc}$ is set to 0. The magnitudes and angles are
in the unit of $eV$'s and degrees, respectively.}
\label{tab:comb_fit}
\end{table}

The results for the fit are given in Table~\ref{tab:comb_fit}.
Again the quality of fit is very good, $\chi^2_{\rm min}/{\rm d.o.f}=1.5/2$.
The weak phase $\ga=(65.5 \pm 14.1)^\circ$ is consistent with
the global fit value by the CKMfitter group~\cite{CKMfitter}.
For $R$, we get $R = 1.79 \pm 2.07$.

In Table~\ref{tab:comb_fit}, the solutions in the first (second) row corresponds to
the black (white) solution in Figure~\ref{fig:CPA}.
For the first solution, the ratio $|C/T|$, given by
\bea
|C/T| = 0.28 \pm 0.15,
\eea
is reduced by about factor 3
compared to (\ref{eq:CoverT}). This agrees with the conventional hierarchy
in (\ref{eq:hierarchy}).
However, as we discussed in section~\ref{sec:B2pipi}, the
ratio $|\wt{C}/\wt{T}|$ should be around 0.7.
The large difference between the ratios $|C/T|$ and $|\wt{C}/\wt{T}|$
can be explained by the constructive interference between $C$ and $P_{uc}$.
As we show in the first solution in Table~\ref{tab:comb_fit},
$C$ and $P_{uc}$ are almost same in sizes, while the difference
between strong phases $\de_{C}$ and $\de_{uc}$ is around $\pi$.
Hence the first solution results in the large enhancement in
$\wt{C} = C - P_{uc}$.
The large $\wt{C}$ can accommodate the ${\cal B}(B^0 \to \pi^0 \pi^0)$.
In Figure~\ref{fig:CPA}, we see that this solution predicts large
positive ${\cal S}_{\rm CP}(B^0 \to K^0 \ol{K^0})$. Numerically we
predict
\bea
{\cal S}_{\rm CP}(B^0 \to K^0 \ol{K^0}) &=& 0.99 \pm 0.02.
\label{eq:SCP1}
\eea
Small errors in (\ref{eq:SCP1}) is because the large errors appearing in
 $r$ are canceled in the numerator and denominator.

The second solution which gives too large ratio $|C/T|=1.34 \pm 0.50$, looks unphysical,
if we take the central value seriously.
This can be tested
by measuring the ${\cal S}_{\rm CP}(B^0 \to K^0 \ol{K^0})$ more precisely. The Figure~\ref{fig:CPA}(b)
shows that this solution predicts large negative
${\cal S}_{\rm CP}(B^0 \to K^0 \ol{K^0})$. The numerical prediction is
\bea
{\cal S}_{\rm CP}(B^0 \to K^0 \ol{K^0}) &=& -0.97 \pm 0.19.
\eea

In~\cite{B2KK2}, the authors could extract the diagrammatic amplitudes from
${\cal B}(B^0 \to K^0 \ol{K^0})$ and a theoretical input
which is free of the endpoint infrared divergences in QCD factorization.
The prediction
\bea
 {\cal S}(B^0 \to K^0 \ol{K^0})= 0.97 \pm 0.02
\eea
is in full accord with the prediction (\ref{eq:SCP1}), reinforing
the conventional SM hierarchy (\ref{eq:hierarchy}). 

\section{Conclusions}
\label{sec:conclusion}
We have considered charmless nonleptonic decay modes, $B \to \pi \pi$ and $B \to K K$.
We find that the large ratio of color-suppressed tree to the color-allowed tree
which is known as the ``$B \to \pi \pi$ puzzle"
is mainly due to the up- and charm-quark penguin $P_{uc}$ contributions which are always
included in the tree diagrams in the form of $\wt{T}=T + P_{uc}$ and $\wt{C}=C - P_{uc}$.
The $P_{uc}$ can be determined with two-fold ambiguity by using
the measured ${\cal B}(B^+ \to K^+ \ol{K^0})$ and ${\cal A}_{\rm CP}(B^+ \to K^+ \ol{K^0})$.
The two-fold ambiguity can be lifted when ${\cal S}_{\rm CP}(B^0 \to K^0 \ol{K^0})$
is measured with more accuracy. Consequently, the bare ratio $|C/T|$ without the
contamination of $P_{uc}$ can be determined.

Conversely, we can predict ${\cal S}_{\rm CP}(B^0 \to K^0 \ol{K^0})$
by imposing the conventional hierarchy to the solutions. For the
theoretically preferred value $|C/T| \sim 0.3$, we get
\bea
  {\cal S}_{\rm CP}(B^0 \to K^0 \ol{K^0}) = 0.99 \pm 0.02.
\eea
If the $B \to \pi\pi$ puzzle disappears in this way, the
$B \to \pi K$ puzzle which is represented by $|C'/T'|=1.6 \pm 0.3$~\cite{B2piK_07} will
become more prominent. Its solution will require other mechanisms in the SM or new physics
beyond the SM.

\vskip1.3cm
\noindent {\bf Acknowledgment}\\
The author thanks C. Yu for carefully reading the manuscript and for useful comments.
The work was supported by
the Basic Research Program of the Korea Science and Engineering
Foundation (KOSEF) under grant No.~R01-2005-000-10089-0.

\end{document}